\documentclass[namedreferences,hyperref,optionalrh,solaromanenum]{spr-sola}
\usepackage{graphicx}
\usepackage{amssymb}  
\usepackage{amsmath}      % useful mathematical symbols
\usepackage{color}    
\usepackage{fixltx2e}       % For color text: \color command

\usepackage{textcomp}     
\usepackage{gensymb}
\usepackage{authblk}  
\usepackage{subfig}

\usepackage{url}        % For breaking URLs easily trough lines
\chardef\us=`\_
\begin{document}
\begin{frontmatter}
\title{A consistency check for the calibration of 5303{\AA} solar coronal emission line observations with Aditya-L1/VELC}
%\title{Radiometric validation of Aditya-L1/VELC}
% using solar disk and Sirius-A observations}
\author[addressref=aff1,corref,email={muthu.priyal@iiap.res.in}]{\inits{V. Muthu Priyal}\fnm{V. Muthu Priyal}}
\author[addressref=aff1]{\fnm{R. Ramesh}}
\author[addressref=aff1]{\fnm{{Jagdev}~\snm{Singh}}}
\author[addressref=aff1]{\fnm{K. Sasikumar}~\snm{Raja}}
\author[addressref=aff1]{\fnm{P. Savarimuthu}}

\address[id=aff1]{Indian Institute of Astrophysics, Koramangala, Bengaluru - 560034}
\runningauthor{Muthu Priyal et al.}
\runningtitle{\textit{Radiometric calibration of VELC}}

\begin{abstract}
We carried out radiometric calibration of the Visible Emission Line Coronagraph (VELC) onboard Aditya-L1 at 5303{\AA} using observations of the Sun itself, a novel approach compared to the calibration using bright stars in the other currently operational space solar coronagraphs. The measured VELC detector count corresponding to the center of Sun's disk is 
${\approx}(1.70{\pm}0.12){\times}10^{8}\, \rm sec^{-1}{\AA}^{-1}$, which relates to Sun's flux at 1\,AU in 5303{\AA}, 
${\approx}3.0456{\times}10^{6}\, \rm {erg\,sec^{-1} cm^{-2} {\AA}^{-1}}$. We verified the above calibration using observations of the bright star Sirius-A whose estimated flux at 1\,AU in 5303{\AA} is
${\approx}1.44{\times}10^{-8}\,\mathrm{erg\,sec^{-1} cm^{-2} \AA^{-1}}$. The expected detector count with VELC in this case is ${\approx}19{\pm}1.4\, \rm sec^{-1}{\AA}^{-1}$,
and the measured count is ${\approx}23{\pm}0.4\, \rm sec^{-1}{\AA}^{-1}$. The reasonable agreement between the expected and observed values for Sirius-A in this initial consistency check after two years of observations with VELC, indicates that 
%VELC’s optical throughput and detector response are performing as designed %in orbit. This validation demonstrates the feasibility of employing bright %stellar sources for in-orbit calibration. It ensures that VELC’s the
measurements of coronal brightness at 5303{\AA} with the VELC and its conversion  into absolute physical units using observations of the Sun disk light are in order. Using the Sirius-A observations we measured the point spread function (PSF) of the VELC in its 5303{\AA} channel also. Its full width at half maximum (FWHM) is 
${\approx}3.8^{\prime\prime}$. 
\end{abstract}

% Keywords
\end{frontmatter}

\section{Introduction}

Ground based observations of the solar corona are inherently limited in both duration and frequency. Total solar eclipses offer only a few minutes of visibility at any given location.
%, requiring careful planning and coordination of expeditions to the path of %totality.
% \citep{habbal2010, boe2020}. 
Ground-based coronagraphs, while capable of continuous observations,
% under ideal conditions, 
%are typically constrained to a few hours of operation per day and 
require exceptionally clear skies. These limitations significantly restrict the ability to monitor dynamic coronal phenomena, e.g. coronal mass ejections (CMEs), over extended periods. To overcome these constraints and enable continuous, 
%high-resolution 
monitoring of the solar corona throughout the year, a space-based coronagraph %with high spectral and spatial resolutions 
is essential. Coronal observations 
%very 
close to the solar limb is another requirement. To address the above, the scientists and engineers at the Indian Institute of Astrophysics (IIA) developed the VELC with help from the Indian Space Research Organisation (ISRO). The VELC, and payloads from other Indian institutions for remote sensing as well as insitu observations of the Sun from the Sun-Earth Lagrangian L1 point were recently launched by ISRO onboard Aditya-L1 \citep{Parate2025}.

A critical aspect of utilizing VELC for quantitative coronal studies is radiometric calibration. This process ensures that the observed detector counts can be converted into absolute physical units of brightness, allowing for better scientific interpretation of the observations. 
However, performing absolute calibration in space presents significant challenges. It is difficult to use traditional light sources used in the laboratory for in-orbit calibration. 
%The other major issues are optical throughput variations, changes in the %detector characteristics, and stray light within the instrument. 
So, bright stellar objects are used.
In the case of the Large Angle Spectrometric Coronagraph (LASCO) onboard the Solar and Heliospheric Observatory (SoHO), observations of stars are used for photometric calibration \citep{Llebaria2006,Morrill2006,Thernisien2006,Gardes2013,Colaninno2015}. Calibration using stars is performed in the METIS coronagraph onboard Solar Orbiter also \citep{DeLeo2023}. For COR1 coronagraph onboard the Solar TErrestrial RElations Observatory (STEREO), observations of Jupiter are used for similar calibration \citep{Thompson2008}.
%Similar calibration scheme for the LASCO-C3 coronagraph is described in 
%\cite{Morrill2006}.}
%astrophysical sources with well-characterized spectral properties are often %employed as secondary calibration standards.
Unlike the above mentioned space coronagraphs, VELC has the capability to independently observe the Sun disk light itself for calibration, 
%does not include an onboard uniform light source for direct calibration. %Nevertheless, it 
%can observe the Sun disk light 
through a Neutral Density (ND) filter and by off-pointing the satellite from the Sun center direction by typically ${\pm}16^{\prime}$. In addition, it can also observe bright star like Sirius-A (without the ND filter), 
by pointing the satellite in its direction. 
%Off-pointing involves aligning the occultor center of the coronagraph with %the target star instead of the usual Sun-centered pointing, allowing stellar %flux measurements in a configuration similar to solar observations.
%In this study, we use Sirius-A, the brightest star in the sky, along with %the solar disk as calibration sources to perform in-orbit radiometric %validation of VELC spectroscopic observations at 5303 {\AA}, as the other %channels were severely affected by noise in orbit. 
By comparing the observed detector counts with the VELC for the above two sources and their reported flux values, we assess the 
%in-orbit performance and 
photometric calibration of VELC using the Sun disk light which is a novel approach for calibration of space solar coronagraph observations.
% from the Sun-Earth Lagrangian L1 location. 
%In addition to validating radiometric response of VELC, this approach %demonstrates the feasibility of using bright stellar sources for in-orbit %calibration of space-based solar coronagraphs also.
%, ensuring reliable measurements of coronal brightness and dynamics.

\section{The Instrument}

\begin{figure}
\centering
\includegraphics[width=0.99\textwidth,clip=]{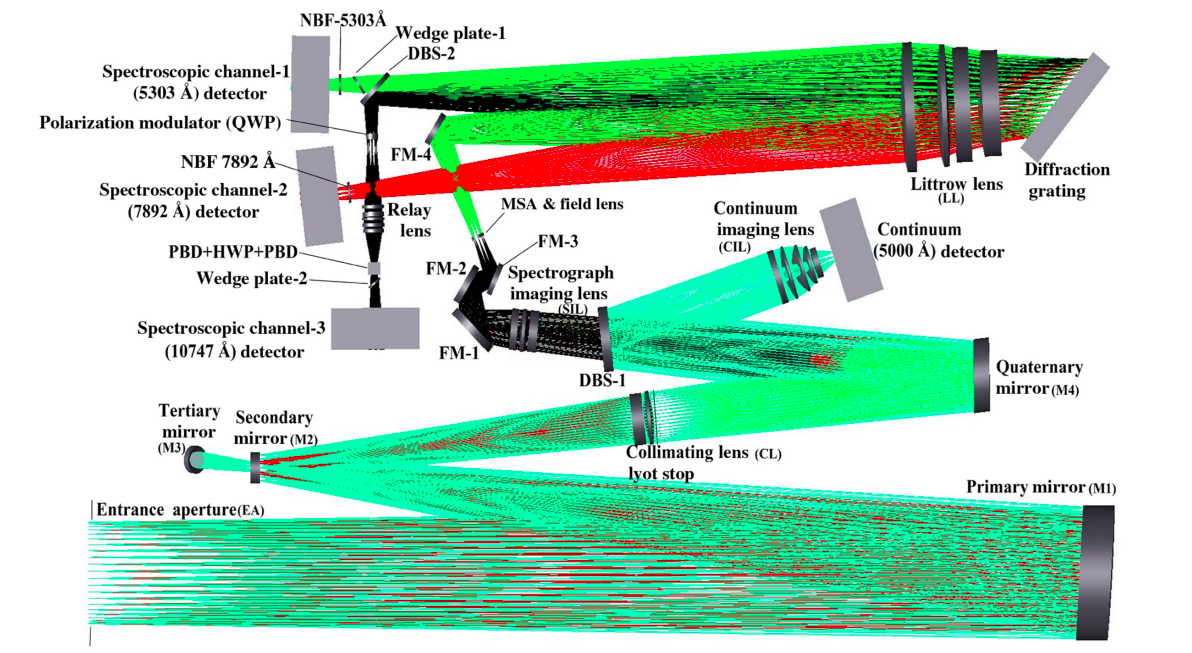} 
\caption{The optical layout of VELC.}
\label{fig:0} 
\end{figure}

The VELC payload onboard ADITYA-L1 is an internally occulted solar coronagraph designed to carry out simultaneous imaging and spectroscopy
%(5000{\AA}) and spectroscopy (5303{\AA}, 7892{\AA}, \& 10747{\AA}) 
observations from close to the limb 
$r{=}1.05R_{\odot}$. 
%The 10747{\AA} channel can be operated in either spectroscopy or %spectropolarimetry mode. 
The field-of-view (FoV) for the imaging channel (continuum) is 1.05\,-\,3.0$R_{\odot}$. 
For the spectral channels, it is 
1.05\,-\,1.5$R_{\odot}$.  
The spectroscopic observations are carried out using four slits simultaneously. 
The continuum and spectroscopic channels have independent narrow-band filters. This helps to avoid overlap of spectra from the different slits in the case of spectral observations.
Figure~\ref{fig:0} shows the optical layout of VELC \citep{Prasad2017,Singh2019}. Sunlight enters the payload via the entrance aperture. The primary mirror forms the image of the Sun disk and corona on the secondary mirror.
Sun disk light and coronal light till $r{=}1.05R_{\odot}$ pass through the central hole of secondary mirror and are reﬂected out of the payload by the tertiary mirror. Coronal light in the 
range $r{=}1.05$\,-\,$3.0R_{\odot}$ is reflected by the annular region of the secondary mirror towards the quaternary mirror which in turn reflects the light towards the dichroic beamsplitter 1 (DBS1).
% which splits the light into two parts. 
The latter reflects the light at wavelengths $<$5100{\AA} towards the continuum imaging lens assembly and narrow band filter (NBF) to form image of the corona in continuum radiation (5000{\AA}) at the CMOS detector. The transmitted light by DBS1 at wavelengths $>$5100{\AA} passes through the imaging lens assembly. Then it is reflected by the mirrors FM1 \& FM2 mounted on a linear scan mechanism (LSM), FM3, and forms image of the corona on the four slits of the grating spectrograph. DBS2 separates the light at wavelengths 10747{\AA} and 5303{\AA}. The combination of Littrow lens, grating and narrow band filters help to record the spectra in three emission lines (5303{\AA}, 7892{\AA}, \& 10747{\AA}) simultaneously using three different detectors.  
%The spectra recorded are analyzed to derive the parameters of the emission line %such as intensity, width, and Doppler velocity at the respective locations of the %corona observed. 
More details of the VELC payload can be found in \cite{Singh2025}. The required pointing of the satellite is achieved using a combination of active pixel Sun sensors
(HAPSS) and star sensors \citep{Parate2025}.

\begin{figure}
\centering
\includegraphics[width=0.99\textwidth,clip=]{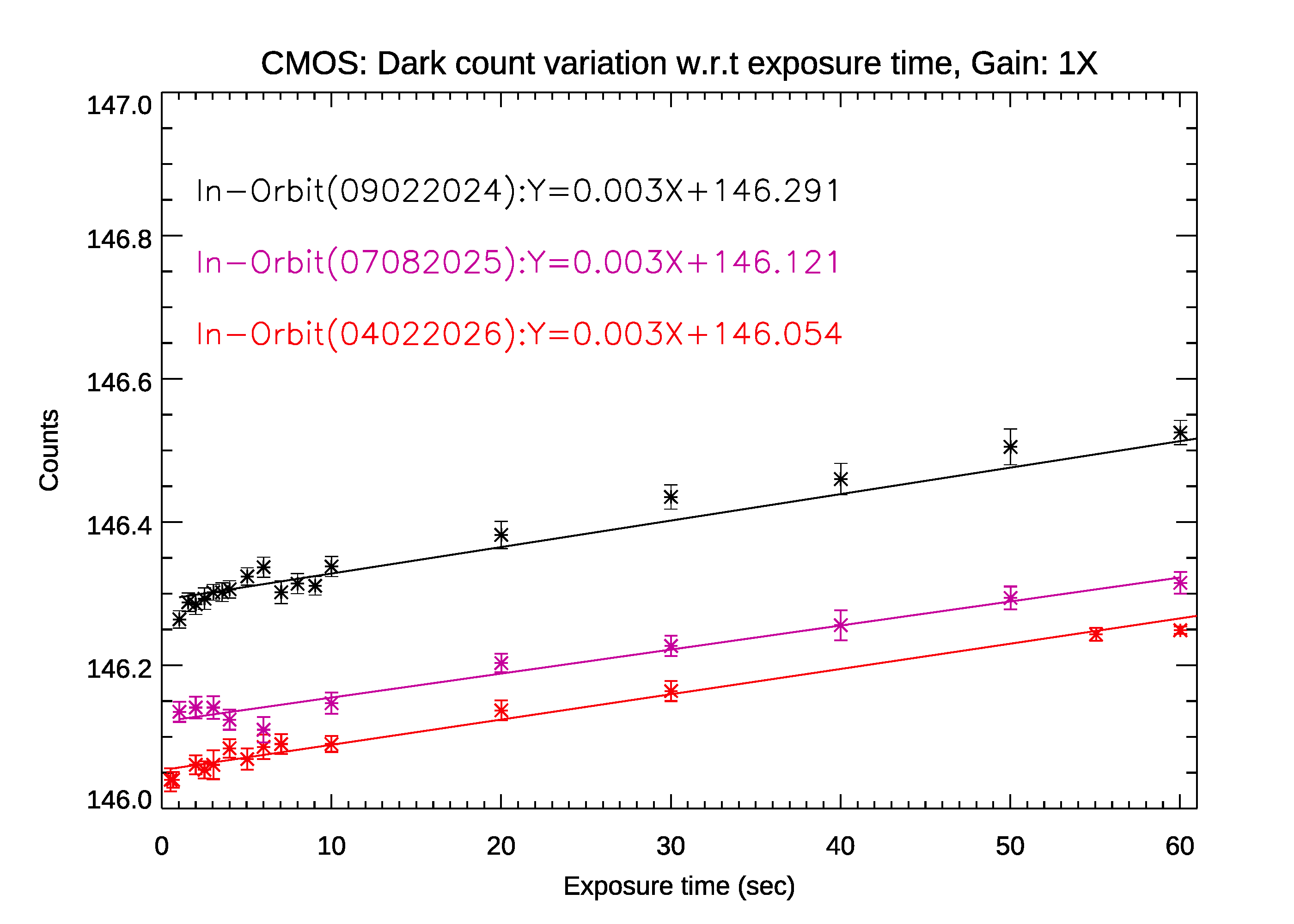} 
\caption{In-orbit `dark' count measurements in the sCMOS detector used in the VELC 5303{\AA} channel, at different epochs (2024 February 9, 2025 August 7, and 2026 February 4). The least squares fits to the data points and their equations are also shown.}
\label{fig:0a} 
\end{figure}

\begin{figure}
\centering
\includegraphics[width=0.99\textwidth,clip=]{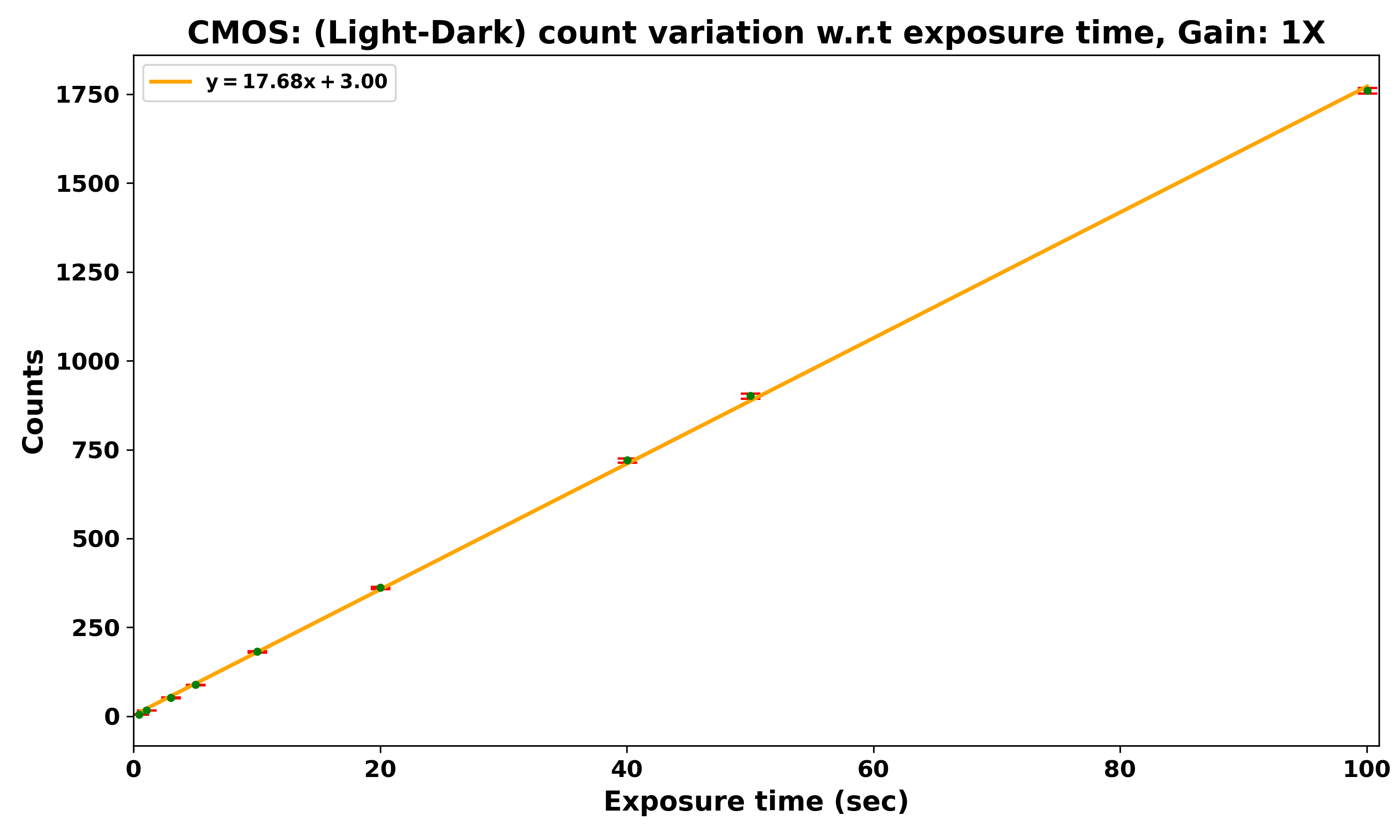} 
\caption{(Light-dark) count measurements with the sCMOS detector used in the VELC 5303{\AA} channel for different exposure times, and uniform illumination. The amplitude of the error bars shown vary from ${\pm}$0.3 counts to ${\pm}$4 counts, from the lowest (444\,msec) to highest (100\,sec) exposure times. The yellow line is the linear least squares fit to the data points.}
% The correlation between the data points is ${\approx}$99\%.}}
\label{fig:0b} 
\end{figure}

For observations in the solar imaging channel at 5000{\AA} and spectroscopy channels at 5303{\AA} \& 7892{\AA}, sCMOS detectors with
spectral response in a range of 4000-7000{\AA} are used. They have the facility to record
the data in low and high gain, simultaneously. The low gains are either 1x or 2x. The corresponding high gains are either
10x or 30x. In the case of  10747{\AA} channel, InGaAs detector with
spectral response in the range of 9000-17000{\AA} is used. The sCMOS and InGaAs detectors are operated at -5$^{\circ}$C and -17$^{\circ}$C, respectively, for optimal performance.  The details about the detectors used in VELC, and their test results are described in \citet{Kumar2024,Mishra2024}. Figure~\ref{fig:0a} shows the consistency in the sCMOS detector `dark' counts measured in-orbit at different epochs, by off-pointing the 
satellite 
%15$^{\circ}$ 
away from the direction of Sun \citep{Singh2025}. 
It remains at nearly the same level as the `ground' measurements. 
Any change as a function of exposure time is also very minimal.  None of the pixels indicated extreme or very low signal with exposure times in the range 0.1\,-\,100 sec. Test results
%, both on-ground and in-orbit, 
indicate that the detector response is linear as a function of the exposure time for uniform illumination (Figure~\ref{fig:0b}). Neither the `dark' frames nor the images obtained with uniform light illumination show any hot/dead pixels.

\begin{figure}
\centering
\includegraphics[width=0.99\textwidth,clip=]{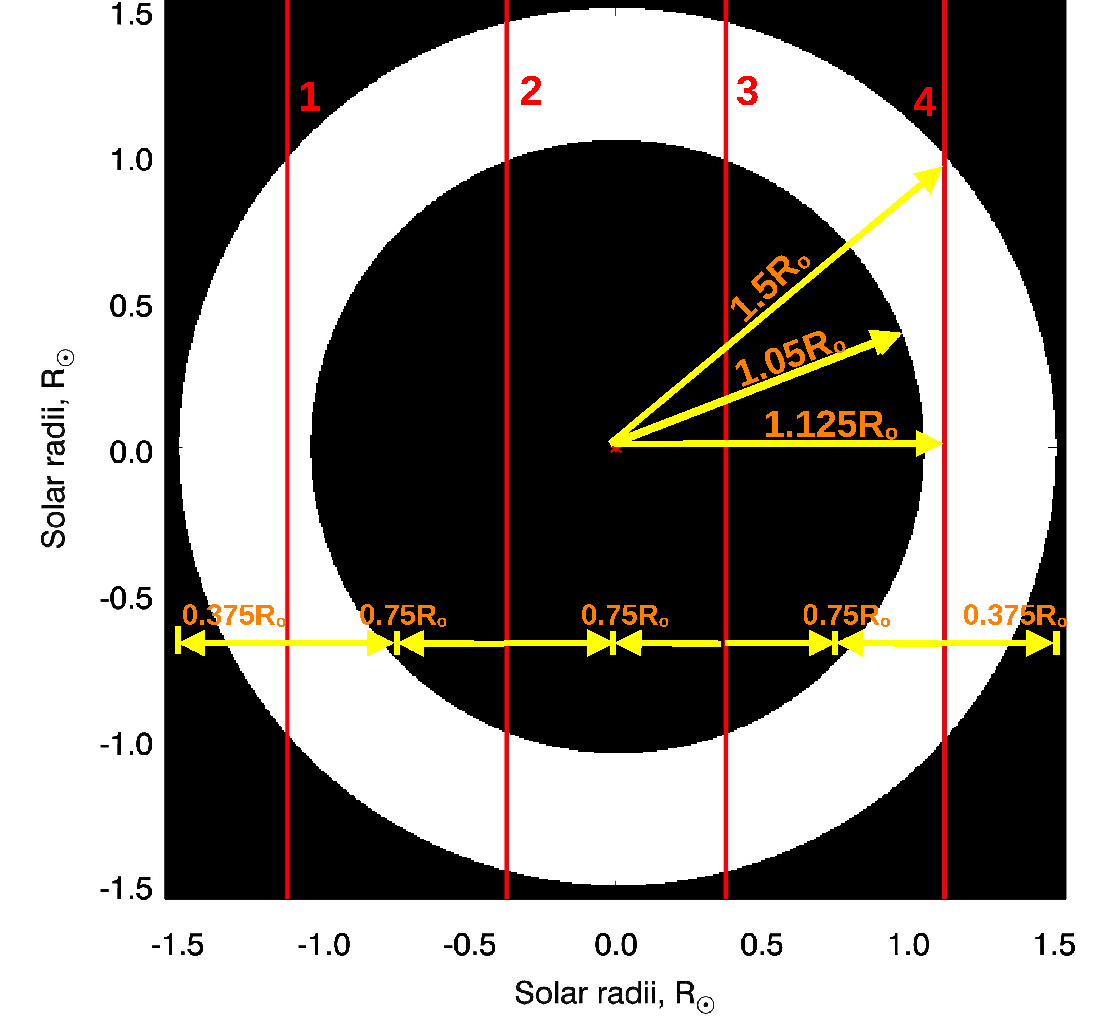} 
\caption{Schematic of the positions of the four slits (represented by the vertical red lines) in the VELC spectroscopy channel. The inner black circle corresponds to the occulter (radius\,=\,$1.05R_{\odot}$). The white circular patch indicates the field-of-view (1.05\,-\,1.5$R_{\odot}$). Solar north and east directions are straight up and to the left, respectively.}
% Right: Intensity image of the solar corona generated using raster scan %data obtained with slit 4 on 16 July 2024. The vertical white line %indicates the FoV of slit 4 when its center observes the corona at 
%$r{\approx}1.05\, \rm R_{\odot}$.}
\label{fig:1} 
\end{figure}

Figure~\ref{fig:1} shows the schematic of the four straight slits in the VELC for spectroscopic observations. While slits 1 \& 4 are located at distance of 
1.125$R_{\odot}$ from the center of the occulter on its either side above the east and west limbs of the Sun, slits 2 \& 3 are located in between. The slits are equi-distant from each other. The spacing between the adjacent slits is 
0.75$R_{\odot}$. The width of each slit is ${\approx}$50${\mu}$m. The height is 
${\approx}$15mm. The spectral and spatial directions of the slit are in the East-West and North-South directions, respectively. The LSM is used to position the image on the slits. The FoV of each slit is
$0.75R_{\odot}{\times}3R_{\odot}$ (solar east-west ${\times}$ solar north-south).  Currently observations in the 5303{\AA} line are routinely carried out \citep{Ramesh2024,Ramesh2025,Muthupriyal2025a,Muthupriyal2025b}. The sCMOS detectors used in the 5303{\AA} channel have 2592${\times}$2192 pixels with 11-bit readout. The plate scale is 1.25$^{\prime\prime}$ per pixel. The pixel size is 6.5${\mu}$m$^{2}$. The spectral dispersion is 28.4m{\AA} per pixel.

\section{Observations and Methodology}

\subsection{Sun disk observation}

Light from the Sun disk center can be used as a standard source for flux calibration, as its brightness varies by only ${\approx}0.2\%$ over a solar cycle \citep{Solanki2013,Frohlich2004,Shapiro2020}. Since VELC is designed to study the solar corona, which is $10^{-6}$ times fainter than the Sun disk light, a 45\,mm aperture space qualified, custom built, reflective ND filter made of synthetic fused silica is kept at the center of the 147\,mm Entrance Aperture (EA) shutter of the VELC. 
Our laboratory measurements indicate that the ND filter has a transmission of 
${\approx}0.64{\times}10^{-4}$ at 5303{\AA} \citep{Singh2025}. 
Note that to characterize the ND filter in the laboratory, first we obtained data without the ND filter for different exposure times by illuminating the EA using 
a quartz tungsten halogen lamp\footnote{https://www.newport.com/p/66295-1KQ-R1}.
%Sun disk light.
% using a Sun tracker, lens arrangement, and an optical fiber cable. 
After subtracting the dark count corresponding to the respective exposure times, the data are normalized to obtain mean count per sec. The same procedure is repeated with the ND filter on the EA. The transmission of the ND filter is calculated as the ratio of the mean count per sec with the ND filter to the mean count per sec without the ND filter. The measured values at different wavelengths (in addition to the 5303{\AA} value mentioned above) are, ${\approx}0.59{\times}10^{-4}$ at 5000{\AA}, 
${\approx}1.58{\times}10^{-4}$ at 7892{\AA}, and ${\approx}2.92{\times}10^{-4}$ at 10747{\AA}. Due to the above characteristics of the ND filter, the Sun disk light passing through it is significantly attenuated,
% by approximately five orders of magnitude, 
preventing saturation of the detector and hence enabling observations of the Sun disk light for calibration purposes in-orbit also. For these measurements, the VELC EA is closed, allowing sunlight to pass only through the ND filter. Observations are carried out by off-pointing the satellite $16^{\prime}$ away from the Sun center, either towards the east or west, so that the disk light will be incident in the VELC field of view,
% in all four directions (east, west, north, and south), 
avoiding the occultor.
% which otherwise is designed to block the Sun disk light. 
%The VELC spectrograph slit was positioned at the edge of the occultor, such %that 
A small portion of the Sun disk, of angular size ${\approx}9.6^{\prime\prime}$ in the East-West direction and ${\pm}1.0{R}_\odot$ in the North-South direction along the central meridian of the Sun, can be observed through either slit 1 or 4 depending on the off-pointing position.

\begin{figure}
\centering
\includegraphics[width=0.99\textwidth,clip=]{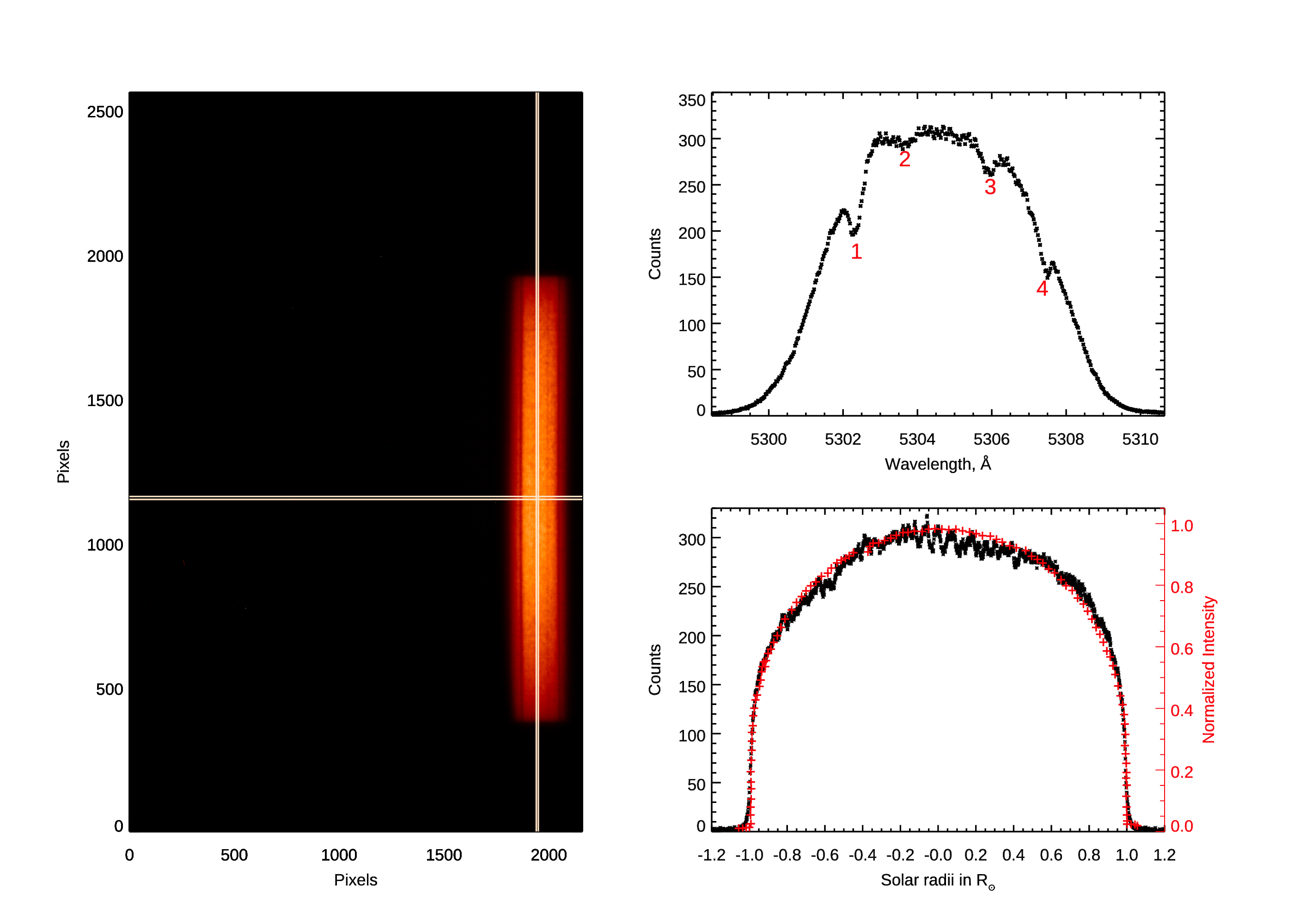} 
\caption{{\bf Left:} Sit and stare observations of the Sun disk light in 5303{\AA} with slit 4 of the VELC on 2024 April 1. The EA shutter is closed and the satellite/occulter center is off-pointed 16$^{\prime}$ away to the east from the direction of Sun center. Spectral and spatial directions are along the horizontal and vertical axis, respectively. 
{\bf Right upper panel:} Spectral profile corresponding to the region near the center of slit indicated by the horizontal lines in the left panel. The numbers 1,2,3 and 4 shown in red colour indicate the absorption lines mentioned in the main text. 
%The detector counts mentioned in the y-axis are in units of per sec per pixel. 
The shape of the profile is due to the narrow band filter (width\,${\approx}$\,6.5{\AA}) used in the VELC. {\bf Right lower panel:} Sun disk profile (in black colour)  corresponding to the region indicated by the vertical lines along the spatial direction of the slit in the left panel. The related VELC detector count values are shown in the left hand side y-axis. For comparison, Sun disk emission at reported in the literature is also shown (in red colour). Its normalized intensity is mentioned in the right hand side y-axis.}
\label{fig:2}
\end{figure}

\begin{figure}
\centering
\includegraphics[width=0.99\textwidth,clip=]{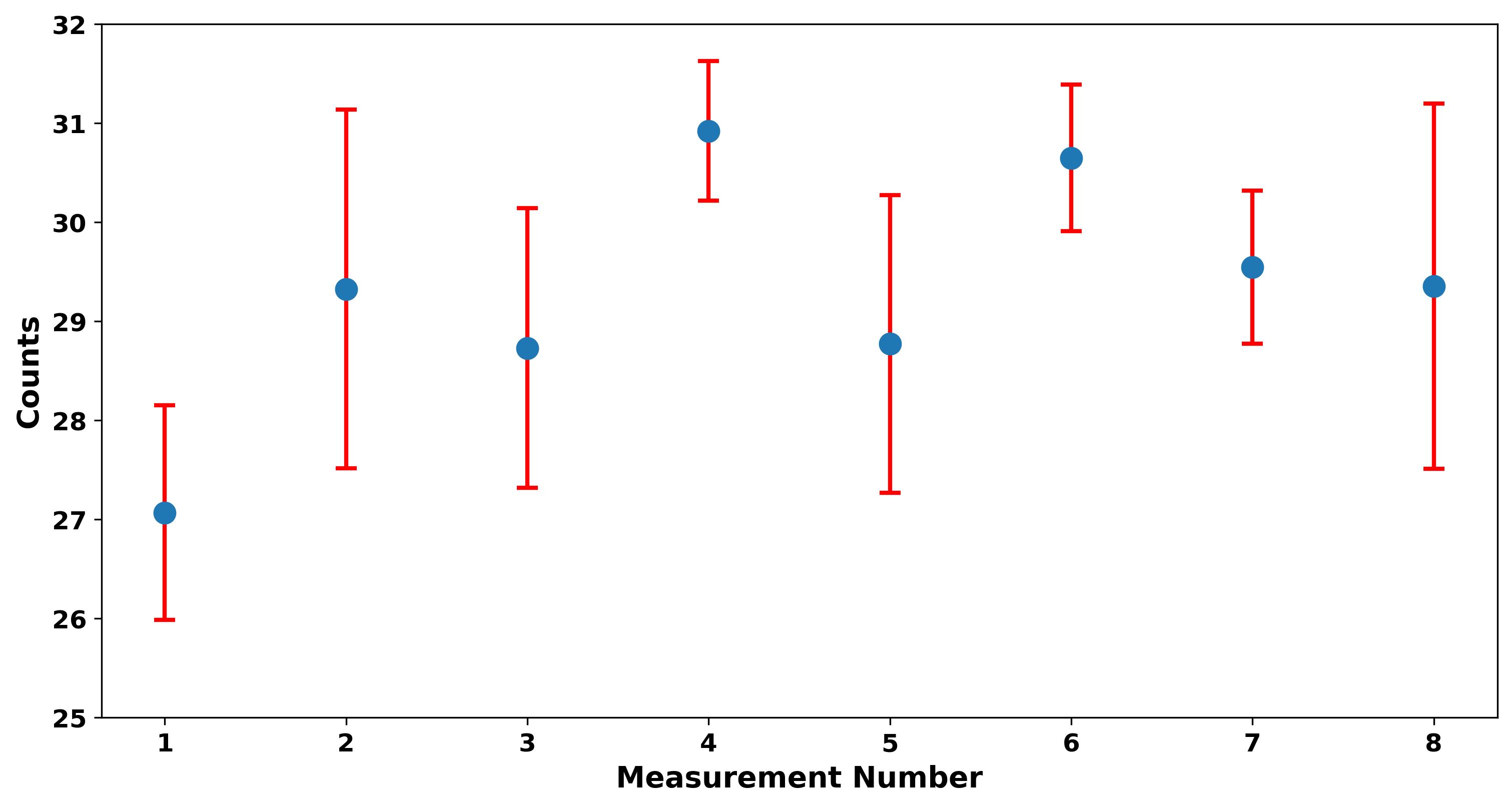} 
\caption{VELC detector counts (per sec, 1x gain) corresponding to the Sun disk center in off-pointing observations and entrance aperture closed condition. The measurements were obtained at multiple epochs between the years 2024 and 2026.}
\label{fig:2a} 
\end{figure}

The left panel of Figure~\ref{fig:2} shows the observed Sun disk spectra, obtained with slit 4.
%at the Sun disk center along the solar N-S direction 
%after dark subtraction. 
after correction for dark
current, curvature of the spectra, ﬂat ﬁeld, and background as
described in \citet{Singh2025}. The upper right panel of Figure~\ref{fig:2} shows the spectral profile at the center of the slit, i.e. disk center. Four absorption lines at wavelengths 5302.3{\AA} (1), 5304.17{\AA} (2), 5305.86{\AA} (3), and 5307.36{\AA} (4) can be seen. They were identified using the solar atlas\footnote{\url{https://nispdata.nso.edu/ftp/pub/atlas/fluxatl/}}.
%The fit to the profile is shown in red and from the fit the filter profile %width is estimated as 7.2{\AA} which is matching with the design value of %7{\AA}. 
%The spectral dispersion is 28.4m{\AA}. 
The exposure time is 5\,sec, and gain is 2x.
% \citep{Singh2022,Muthupriyal2024}. 
The measured detector count in the continuum portion of the spectrum adjacent to the 5304.17{\AA} absorption line, after conversion for exposure time and gain, is 
${\approx}29{\pm}1\, \rm sec^{-1}$ for 1x gain. The above value corresponds to the center of Sun's disk. Similar measurements obtained from in-orbit observations at several epochs over the last two years are in the range ${\approx}$27\,-\,31 counts $\rm sec^{-1}$
(Figure~\ref{fig:2a}).
The variations w.r.t the mean value (${\approx}$29 counts $\rm sec^{-1}$) are 
${\lesssim}$7\%.
%${\approx}30{\pm}1\, \rm sec^{-1}$, ${\approx}27{\pm}1\, \rm sec^{-1}$, and 
%${\approx}31{\pm}1\, \rm sec^{-1}$. 
%, primarily due to satellite pointing instabilities. 
%, due to the uncertainity of ${\pm}$2 counts, 
% pixel^{-1}$  
Taking into consideration the spectral dispersion of 28.4m{\AA} per pixel (Section 2), the 
above mean detector count (${\approx}29{\pm}2\, \rm sec^{-1}$) is  
${\approx}1021{\pm}70\,\rm {sec}^{-1}{\AA}^{-1}$. The lower right panel of Figure~\ref{fig:2} shows the Sun disk profile (in black colour) observed with the VELC, along the spatial direction of the slit. It matches nicely with the typical Sun disk emission profile, shown in red colour (see, e.g. \citealp{Petro1984}). There is good correspondence between the limb darkening also in the two profiles.
%\, (29 \times 1000 / 28.4)$.
% (after conversion to units of per sec per {\AA}).
%It should be noted that this value corresponds to the integrated count over the solar disk of size $9.6^{\prime\prime}$ x $\pm 1.0\mathrm{R}_\odot$.

\begin{figure}
\centering
\includegraphics[width=0.99\textwidth,clip=]{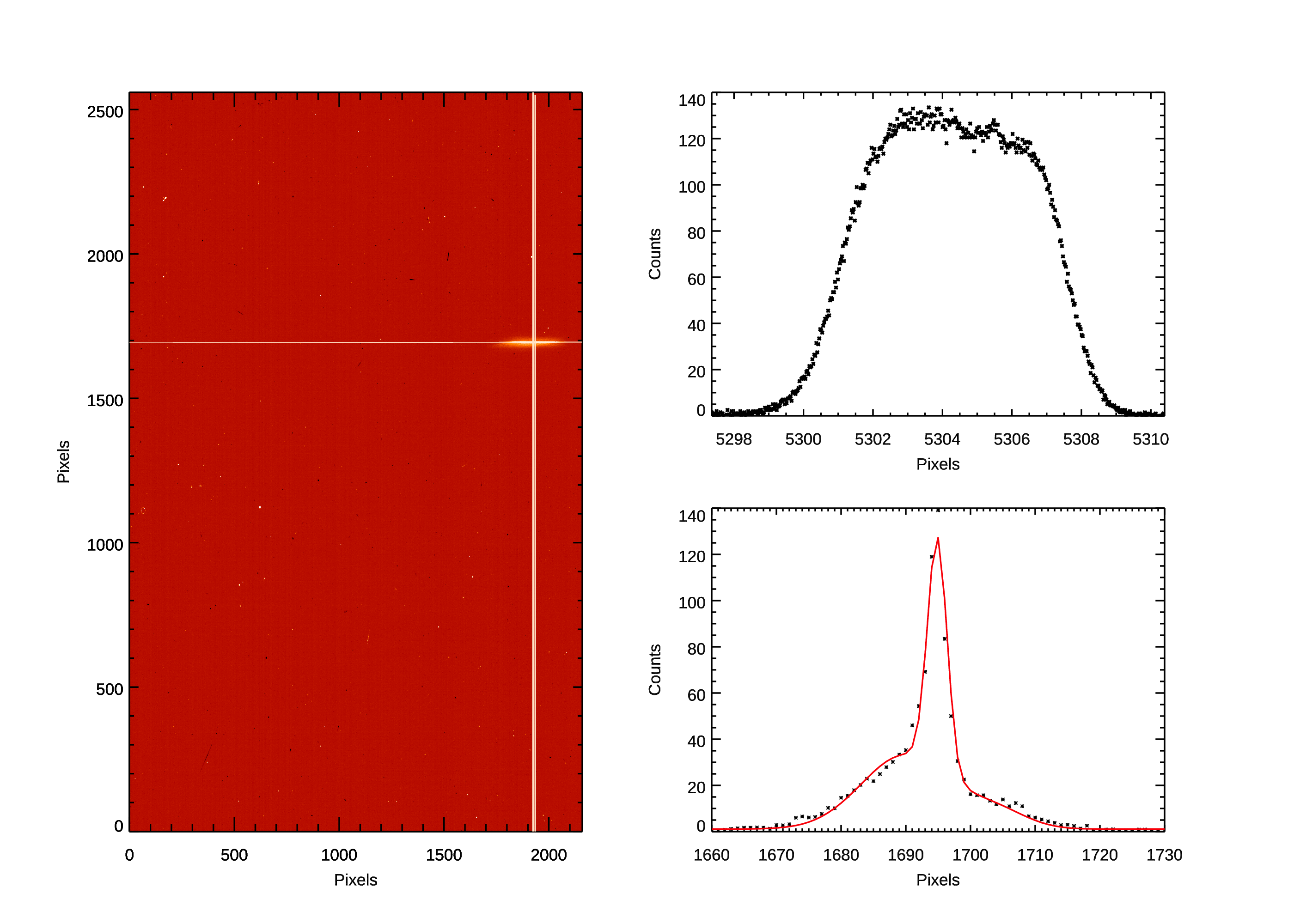} 
\caption{{\bf Left:} Sit and stare observations of Sirius-A in 5303{\AA} with slit 4 of the VELC on 2024 December 11, similar to Figure~\ref{fig:2}. The EA shutter is open and the satellite is off-pointed 19$^{\prime}$ away from the direction of Sirius-A. The bright emission near the intersection of the vertical and horizontal lines is the spectrum from Sirius-A. {\bf Right upper panel:} The spectral profile corresponding to the region indicated by the horizontal lines in the left panel. {\bf Right lower panel:} Profile corresponding to the region by the vertical lines along the spatial direction of the slit showing the PSF of VELC 5303{\AA} channel. Its FWHM is ${\approx}$3 pixels. The broadening of the profile closer to its base is likely due to spacecraft pointing instabilities.}
% which is ${\sim}{\pm}25^{\prime\prime}$ (see, e.g. \citealp{Singh2025}).}
\label{fig:3}
\end{figure}

\subsection{Sirius-A observation}

Observations of Sirius-A in the VELC spectral channels are carried out by off-pointing the satellite ($135^{\circ}$, the Sun-Sirius angular separation during the epoch of observations reported in this work) to align the VELC occulter center in the direction of the source. Unlike the Sun disk observations in Section 3.1, the VELC EA is kept open for observations of Sirius-A. Since the latter is a point source, it is completely occulted in the above mentioned pointing position. 
%Light from Sirius-A will pass through the hole in the secondary mirror and   %reflected outisde the payload by the tertiary mirror (Section 2). 
%Subsequently, 
The satellite is off-pointed 
%further off-pointed 
by $\pm 19^{\prime}$ so that light from Sirius-A is observed in the VELC FoV. 
%The observations were conducted in raster scan mode \citep{singh2025}. In %each raster scan step, one frame was acquired with an exposure time of 100 %sec using 2X gain mode. The Linear Scan Mechanism (LSM) step size was %approximately 20 ${\mu}$m. A total of 95 steps were carried out, producing %95 spectra in a single raster scan to cover the entire VELC field of view. %The total observation time for each pointing position was approximately 158 %min.
%From each spectrum, the average count across the filter profile was %estimated at each spatial position along the slit. The raster scan image was %then generated by stacking the average count for all 95 spectra.
The left panel of Figure~\ref{fig:3} shows the spectrum of the light from Sirius-A 
%at a single step of the raster scan 
at 5303{\AA} observed with slit 4,
% with the source offset by $+19^{\prime}$, 
after subtraction of the spectrum obtained when Sirius-A was near the occulter center. 
%Being a point source, 
%Light from Sirius-A passed through the hole in the secondary mirror and was %reflected outisde the payload by the tertiary mirror 
%(Section 2). 
The upper right panel of Figure~\ref{fig:3} shows the
spectral profile of the light due to Sirius-A in the left panel. The exposure time is 100\,sec, and gain is 2x.
%The detector counts are in units of per sec per pixel.
%Its estimated width from the fit is 7.2{\AA} which closely matches with the %design value of 7{\AA} \citep{singh2025}.
% intensity profile at one spatial position across the Sirius. The numerous %bright (hot) pixels seen across the detector are attributable to the long %exposure time (100 seconds), which can produce hot pixels. 
The measured detector count, after conversion for exposure time and gain, is 
${\approx}0.65{\pm}0.01\, \rm sec^{-1}$ for 1x gain. Considering the spectral dispersion per 
% pixel^{-1}$ 
pixel (Section 2), the above count value becomes 
${\approx}22.9{\pm}0.4\,\rm {sec}^{-1}{\AA}^{-1}$.
%\, (0.03 \times 1000 / 28.4)$.
%From the continuum portion of the spectrum, we measure a Sirius-A count rate %of 21 counts per second per {\AA} in 1X gain.
%Distortions in the raster scan image, both in the spatial and scan directions, are due to spacecraft stability. The integrated count for Sirius-A was measured to be 4000 counts for 100 sec with 2X gain. Therefore, the integrated count for Sirius-A per second for 1X gain was 20 counts. 
We measured the PSF in the 5303{\AA} channel of VELC from the 
%full width at half maximum of the 
Sirius-A profile along the spatial direction (lower right panel of Figure~\ref{fig:3}). Its FWHM is 3 pixels. In angular scale, it is $3{\times}1.25^{\prime\prime}{\approx}3.8^{\prime\prime}$.
 %nearly twice the plate scale per pixel, i.e. $2{\times}1.25^{\prime\prime}$ (see, %e.g. \citealp{Brueckner1995}).

\section{Results and Discussions}

The primary objective of the observations mentioned in Sections 3.1\,\&\,3.2 is to establish radiometric calibration for the 5303{\AA} observing channel of the VELC using light from the Sun disk (viewed through a ND filter with EA closed) by comparing it with observations of 
Sirius-A (with EA open). 
%Both are stable sources of known brightness.
%These calibration targets allow us to evaluate the instrument’s response %over a wide dynamic range, from the extremely bright extended solar disk to %a point-like stellar source. 
%By comparing the measured detector counts from both the targets (after %applying appropriate corrections for attenuation, aperture size, exposure %settings, etc.), we can derive the instrument’s effective throughput and %assess its suitability for precise coronal brightness measurements. 
%We carried out observations with VELC for two calibration targets — Sirius-%A, a bright and photometrically stable star that serves as an point source %for VELC, and the solar disk, observed through a Neutral Density (ND) %filter. 
The measured detector count corresponding to the center of Sun's disk   
%(with the VELC EA shutter closed and through the ND filter) 
is ${\approx}1021{\pm}70\, \rm sec^{-1}{\AA}^{-1}$ (Section 3.1). For Sirius-A (VELC EA shutter open and without ND filter), the detector count is 
${\approx}22.9{\pm}0.4\, \rm sec^{-1}{\AA}^{-1}$ (Section 3.2). 
%The corresponding integrated count values are 
%${\approx}28663{\times}10^{4}\, \rm sec^{-1}{\AA}^{-1}$, and 
%${\approx}14943\, \rm sec^{-1}{\AA}^{-1}$, respectively.
%for a total exposure of 2000 seconds in 2X gain mode.
To compare the above mentioned counts from the two sources, it is necessary to normalize the differences in detector settings, attenuation due to the ND filter, aperture size used for the Sun disk measurements, etc. between the two observations. The Sun disk measurements are obtained through the ND filter, so the actual detector count is estimated by applying the ND filter transmission factor and correcting for the ratio of the ND filter aperture to the VELC EA. With the inclusion of the above factors, the detector count corresponding to the light from the center of Sun's disk 
% the center of the Sun disk 
is, $[(1021{\pm}70){\times}(147/45)^{2}]\,/\,[0.64{\times}10^{-4}]\,{\approx}\,(1.70{\pm}0.12){\times}10^{8}\,\rm{sec}^{-1}\,{\AA}^{-1}$.
%$[(1021{\pm}35){\times}(147/45)^{2}]\,/\,[0.64{\times}10^{-4}]\,{\approx}\,%(1.70{\pm}0.06){\times}10^{8}\,\rm{sec}^{-1}\,{\AA}^{-1}$.
%In the same way, we need to compensate for the Earth-Sirius distance, Sun-earth distance and the Sirius and Sun disk size for meaningful comparison. Hence, the measured Sirius count is multiplied by the above factor to compensate for the angular distance between Sirius, Sun and Earth. Thus, the estimated Sirius count is  $5.8{\times}10^{12}$. Then the ratio of the Sirius and Sun disk count measured by VELC is found to be 5.6.
%The Sun-Earth distance is about 150 million km, whereas Sirius-Earth distance is about $8.147{\times}10^{13}$ km. Therefore, brightness of Sirius-A at 1AU will increase by a factor using inverse-square law is $2.94{\times}10^{11}$. 

%To validate the above, Sun and Sirius-A flux values at 5303{\AA} are considered. 
Reports indicate that the flux from Sun at 5303{\AA}, 
%For Sun, the values 
near 1\,AU and Earth
%{\url{https://nispdata.nso.edu/ftp/pub/atlas/fluxatl/}} 
are $3.0456{\times}10^{6}\, \rm {erg\,sec^{-1} cm^{-2} {\AA}^{-1}}$ and 65\,$\mathrm{erg\,sec^{-1}cm^{-2}\AA^{-1}}$, respectively\footnotemark[1]. For Sirius-A, it is $1.44{\times}10^{-8}\,\mathrm{erg\,s^{-1} cm^{-2} \AA^{-1}}$ at Earth \citep{Bell1981,Bohlin2014}. Our objective is to estimate the expected detector count for Sirius-A from the measured VELC detector count corresponding to Sun disk center, at the Lagrangian point (L1). Considering that the Earth-L1 distance is small compared to the distances to Sun and Sirius-A from Earth, we assume that the flux of Sun and Sirius-A at L1 are nearly same as their corresponding values at Earth.
%So, we need to account for the L1\,-\,Sirius distance, the L1\,-\,Sun %distance, the Sirius\,-\,Sun distance, and the angular sizes of Sun 
%($1920^{\prime\prime}$) and Sirius-A ($0.006^{\prime\prime}$). 
%The flux of Sun at 1\,AU, considering the above size difference between Sun %and Sirius-A is, 
%($3.0456{\times}10^{6}$)${\star}$(0.006/1920)\,${\approx}$\,
%\(\dfrac{ \text{Sun flux at 1\,AU} \times 0.006 }{ 32 \times 60 }\). 
%The Sun flux value obtained is 
%$9.518\, \rm {erg\,s^{-1} cm^{-2} {\AA}^{-1}}$.
%The flux of Sirius-A at 1\,AU is $1.45{\times}10^{-8}\,\mathrm{erg\,s^{-1} %cm^{-2} \AA^{-1}}$, nearly same as the value at Earth mentioned above. 
The extent of illumination due to the Sun disk (Fig.~\ref{fig:2}) and Sirius (Fig.~\ref{fig:3}) in the spatial direction are ${\approx}$1536 \& 3 pixels, respectively. 
These correspond to the size of the Sun disk ($1536{\times}1.25^{\prime\prime}{=}1920^{\prime\prime}$), and the FWHM of the PSF 
%($3{\times}1.25^{\prime\prime}{=}3.75^{\prime\prime}$) 
in the case of Sirius-A. 
%In the spectral direction, it is the same for both the sources.
The expected detector count for Sirius-A at L1, derived from the Sun flux, is 
\,=\,[(measured Sun disk center detector count at L1)\,${\times}$\,(Sirius-A flux at L1)]\,/\,[(Sun flux at L1)\,${\times}$\,(3/1536)]
%. The value obtained is 
${\approx}19{\pm}1.4$ counts\, $\rm sec^{-1}{\AA}^{-1}$.
%found to be 
%${\approx}$11 using Equation~\ref{eq1}:
%\begin{equation}
%\label{eq1}
%\text{Expected count} =
%\frac{\text{Measured Sun disk count at L1} \times \text{Sirius-A flux at 1 %AU}}
%{\text{Sun flux at 1 AU}}
%\end{equation}
This result is ${\approx}$4 counts less than the measured value (Section 3.2). 
However, considering that the exposure time used for Sirius-A observations is 100\,sec (Section 3.2) for which the expected error in the (light-dark) measurements is ${\pm}$4 counts (Figure~\ref{fig:0b}), the above difference of ${\approx}$4 counts is reasonable. So we can assume that the VELC is functioning as as expected.
%This result is in {\bf reasonable} agreement with the measured Sirius-A count of 
%${\approx}23{\pm}0.4\, \rm sec^{-1}{\AA}^{-1}$ at L1 (Section 3.2), indicating that VELC is %functioning as expected. 
Furthermore, the above agreement between the expected and measured counts for Sirius-A demonstrates that the 
%absolute 
radiometric response of VELC is stable in orbit, validating the use of Sun disk light observations to calibrate the brightness measurements of the solar corona. We would like to add here that the above mentioned difference of ${\approx}$4 counts in the Sirius-A observations corresponds to an error of (4/23)${\approx}$17\%, which is larger than the 7\% variation in the detector counts in the Sun disk observations (Section 3.1).

\section{Summary and Conclusions}
We carried out radiometric calibration of the Visible Emission Line Coronagraph (VELC) at 5303{\AA} using two stable reference sources: light from the Sun disk, observed through a ND filter, and the bright star Sirius-A. 
%The Sun provides a uniform, extended high-brightness source, while Sirius-A %serves as a point source with a well-known spectral energy distribution.
The measured detector count due to light from the Sun disk center is corrected for the ND filter transmission and aperture size, yielding an estimated total of $(1.70{\pm}0.12){\times}10^{8}\, \rm sec^{-1}{\AA}^{-1}$ counts. For Sirius-A, the expected detector count derived from observations of the Sun disk and the flux values of Sun and Sirius-A at 1\,AU, is 
${\approx}19{\pm}1.4$ counts\, $\rm sec^{-1}{\AA}^{-1}$. Compared to this, the measured 
count with VELC is 
${\approx}23{\pm}0.4\, \rm sec^{-1}{\AA}^{-1}$.
The reasonable agreement between the measured and expected Sirius-A detector counts in this initial consistency check after two years of observations with VELC,
taking into consideration the large exposure time required for observations of Sirius-A (Section 4),
%confirms that VELC’s optical throughput 
%This consistency 
validates the radiometric calibration of VELC, and demonstrates the feasibility of using Sun disk light for the in-orbit calibration of space-based solar coronagraphs like VELC where it is difficult to observe the stars along with the solar images in the same FoV similar to LASCO-C2 (see, e.g \citealp{Colaninno2015}).
% also. Such calibrations are essential for %ensuring accurate and 
%reliable measurements of coronal brightness and verification of the payload  %performance throughout the mission lifetime.
The advantage with flux calibration using the Sun disk light observed via a ND filter is that the related data can be obtained for nearly the same exposure times used in the coronal observations. Propagation of uncertainities from Sirius-A to Sun observations during the process of calibration, for e.g. the errors associated with the requirement of comparatively larger exposure time for Sirius-A as described in Section 4, are not there due to this. In addition to the flux calibration, we measured the spatial resolution in the VELC 5303{\AA} channel also using the Sirius-A observations. It is ${\approx}3.8^{\prime\prime}$.

\section*{Acknowledgments}
Aditya-L1 is an observatory class mission, funded and operated by ISRO.
We thank the scientists and engineers in IIA and ISRO for their contributions to the VELC payload. C. Kathiravan is thanked for discussions on the flux calculations. We are thankful to the referee for kind and insightful comments which helped to present our results more clearly.
% to reach the present state. 

\section*{Declarations}
No funding was received for conducting this study.

\bibliographystyle{spr-mp-sola}

\bibliography{reference.bib}  

%\end{article} 

\end{document}